\documentclass[aps,floatfix,pra,showpacs,12pt]{revtex4-1}
\usepackage{amssymb,amsfonts,mathrsfs}
\usepackage{graphicx,bm}
\usepackage[T1]{fontenc}
\usepackage{color}
\usepackage{xcolor}
\usepackage{hyperref}

\newcommand{\ket}[1]{|#1\rangle}

\newcommand{\bb}[1]{\left( #1 \right)}

\newcommand{\ebefore}{E_{\it current}}
\newcommand{\eafter}{E_{\it candidate}}

\newcommand{\tchar}{T_{\rm p}}
\newcommand{\oppsi}{\hat{\Psi}}
\newcommand{\oppsis}{\hat{\Psi}^{\dagger}}

\begin{document}

\title{Fock State Sampling Method -- Characteristic temperature of maximal fluctuations for interacting bosons in box potentials}

\author{M.~B.~Kruk}
\affiliation{
  Center for Theoretical Physics, Polish Academy of Sciences, Al. Lotnik\'{o}w 32/46, 02-668 Warsaw, Poland
}
\affiliation{Institute of Physics, Polish Academy of Sciences, Al. Lotnik\'{o}w 32/46, 02-668 Warsaw, Poland}

\author{T.~Vibel}
\affiliation{                    
  Center for Complex Quantum Systems, Department of Physics and Astronomy, Aarhus University, Ny Munkegade 120, DK-8000 Aarhus C, Denmark
}
\author{J.~Arlt}
\affiliation{                    
  Center for Complex Quantum Systems, Department of Physics and Astronomy, Aarhus University, Ny Munkegade 120, DK-8000 Aarhus C, Denmark
}

\author{P.~Kulik}
\affiliation{                    
  Center for Theoretical Physics, Polish Academy of Sciences, Al. Lotnik\'{o}w 32/46, 02-668 Warsaw, Poland
}

\author{K.~Paw{\l}owski}
\affiliation{                    
  Center for Theoretical Physics, Polish Academy of Sciences, Al. Lotnik\'{o}w 32/46, 02-668 Warsaw, Poland
}
\email{pawlowski@cft.edu.pl}

\author{K.~Rz\k{a}żewski}
\affiliation{                    
  Center for Theoretical Physics, Polish Academy of Sciences, Al. Lotnik\'{o}w 32/46, 02-668 Warsaw, Poland
}
\date{\today}


\begin{abstract}
We study the statistical properties of a gas of interacting bosons trapped in a box potential in two and three dimensions. Our primary focus is the characteristic temperature $\tchar$, i.e. the temperature at which the fluctuations of the number of condensed atoms (or, in 2D, the number of motionless atoms) is maximal. Using the Fock State Sampling method, we show that $\tchar$ increases due to interaction. In 3D, this temperature converges to the critical temperature in the thermodynamic limit. In 2D we show the general applicability of the method by obtaining a generalized dependence of the characteristic temperature on the interaction strength. Finally, we discuss the experimental conditions necessary for the verification of our theoretical predictions.
\end{abstract}

\maketitle

\section{Introduction}

The statistical properties of interacting ultracold gases of bosonic atoms and in particular of Bose-Einstein condensates remain a considerable challenge of current interest. While the statistical properties of non-interacting gases are well described by a number of methods, a soluble model for interacting bosons exists only in one dimension. In two and three dimensions, reliable results are only available for weakly interacting gases  at low temperatures within the Bogoliubov approximation. Hence, the dependence of the critical temperature on interaction remains a challenging issue.

Over the years a large number of mutually exclusive predictions for the change of the critical temperature due to interactions were made \cite{Lee1957, Grueter1997, Holzmann1999Aug, holzman1999, Baym1999Aug, Wilkens2000Oct, Arnold2000, Baym2000Jan, Cruz2001, Kashurnikov2001, Andersen2004}. Practically all of them dealt with a gas trapped in a three-dimensional cubic box potential. The conflicting results are summarized  in Table~\ref{tab:corrections}. Note, that even the sign of the correction was uncertain initially. Later the  consensus emerged, that in the thermodynamic limit, the shift to the critical temperature is $\Delta T_c \approx 1.3\, \bb{a\rho^{1/3}}$, where $\rho$ is the gas density and $a$ is the s-wave scattering length. 
\begin{table}
\begin{tabular}{| p{0.5\textwidth} || p{0.1\textwidth}p{0.2\textwidth}| }
\hline
Authors & ~&Coefficient $c$ \\
\hline
 Grueter {\it et al.} (1997) \cite{Grueter1997} & ~&$0.34\pm0.06$  \\ 
 Holzmann {\it et al.} (1999) \cite{Holzmann1999Aug} & ~&$0.7$ \\ 
 Holzmann {\it et al.} (1999) \cite{holzman1999} & ~&$2.3\pm0.25$ \\
  Baym {\it et al.} (1999) \cite{Baym1999Aug}  & ~&$2.9$ \\
 Wilkens {\it et al.} (2000) \cite{Wilkens2000Oct} & ~&$-0.93$ \\ 
 Arnold {\it et al.} (2000) \cite{Arnold2000} & ~&$1.71$ \\ 
 Baym {\it et al.} (2000) \cite{Baym2000Jan} & ~&$2.33$ \\
 F. F. de Souza Cruz {\it et al.} (2001) ~\cite{Cruz2001}  & ~&$3.059$\\
  Kashurnikov {\it et al.} (2001) \cite{Kashurnikov2001}  & ~&$1.29\pm0.05 $\\
 M. J. Davis and S. A. Morgan (2003)~\cite{Davis2003Nov}   & ~&$1.3\pm0.4 $\\
Kwangsik Nho and D. P. Landau (2004)~\cite{kwangsik2004}   & ~&$1.32\pm0.14$ \\
S. Watabe and Y. Ohashi
(2013)~\cite{Watabe2013}   & ~&$1$ to $6.7$ \\
\hline
\end{tabular}
\caption{\label{tab:corrections}
Coefficient $c$ of the shift of the critical temperature  $\Delta T_c/T_c = c a \bb{\rho}^{1/3}$ obtained by various analytic and numerical methods. See also the review by  J.~Andersen~\cite{Andersen2004}. In reference~\cite{Watabe2013} four different many-body methods were used yielding results in the range $c\in[1,\;6.7]$.}
\end{table}

During the struggle to compute the shift of the critical temperature a number of theoretical methods were used (see the review~\cite{Andersen2004}). The correct result was eventually obtained by using the classical field approximation (CFA)~\cite{Kashurnikov2001}. The CFA method itself suffers a cut-off problem, which was cleverly overcome in Ref.~\cite{Kashurnikov2001}. Recently, we proposed yet another method based on a direct quantum description of the system and the definitions of the statistical ensembles. The method, called the Fock-State sampling~(FSS) method~\cite{Kruk2022Sep} is presented in Sec.~\ref{sec:method}.

Most BEC experiments to date are performed with harmonic traps. However, recently Bose-Einstein condensates were created in nearly perfect box potentials~\cite{Gaunt2013}. Nonetheless, an experimental verification of the theoretical prediction remains challenging. One of the problems is due to the fact, that experiments are performed with a finite number of atoms and for such a system there is no unique way of determining the critical temperature. Namely, for a finite size system, the number of condensed atoms is an analytic function of temperature, and thus there is not a definite value of temperature beyond which the number of condensed atoms is strictly zero. The remedy for this difficulty proposed in Ref.~\cite{Idziaszek2003}, is to study the temperature of maximal variance of the number of condensed atoms instead of the critical temperature. The temperature of the maximal variance tends to the critical temperature in the thermodynamic limit~\cite{Idziaszek2003}, and it is well-defined for finite size systems, which makes it applicable to gases exhibiting only quasi-condensation.

Experimentally, it is more demanding to measure the fluctuations of the condensate atom number than the mean of this number. However, the experimental difficulties were recently overcome due to a stabilization technique of the evaporation process~\cite{Gajdacz2016}, allowing for a measurement of the fluctuations~\cite{Kristensen2019}. Furthermore, it was shown, that the canonical ensemble fails to describe the experimental situation, and one must invoke the microcanonical one~\cite{Christensen2021}.
These experiments directly measure the temperature of the maximal fluctuations rather than the temperature at which the condensate vanishes.

It is the purpose of this paper to discuss the interaction-induced shifts of the temperature of maximal fluctuations, which is referred to as the characteristic temperature $\tchar$. Based on the FSS method we provide, to our knowledge for the first time, results for a bosonic gas in a box potential in the microcanonical ensemble.

The paper is organized as follows. In Sec.~\ref{sec:method} we briefly review the FSS method. Section~\ref{sec:results3d} applies the method to a gas trapped in the three-dimensional box potential both in the canonical and microcanonical ensembles. In Sec.~\ref{sec:results2d} the case of the two-dimensional box potential is discussed. Note, that there is no phase transition and no critical temperature in this case, and nonetheless the characteristic temperature can be defined. Section~\ref{sec:conclusions} concludes the discussion and provides an outlook on future experiments.

\section{Fock State Sampling Method\label{sec:method}}
We consider $N$ bosonic atoms trapped in a box potential with periodic boundary conditions and interacting via short-range interaction potential. The Hamiltonian of the system is
    \begin{equation}
        \hat{H} = -\frac{\hbar^2}{2m}\int\,{\rm d}^d r\, \oppsis (\bm{r}) {\nabla}^2 \oppsi(\bm{r}) + \frac{g_{\rm d}}{2}\int\,{\rm d}^d r\, \oppsis(\bm{r}) \oppsis (\bm{r}) \oppsi(\bm{r}) \oppsi(\bm{r}),
       \label{eq:hamiltonian}
    \end{equation}
where $\oppsi(\bm{r})$ is a bosonic annihilation operator, $m$ is a mass, and $g_{d}>0$ is a coupling constant related to short-range interactions. In the Sec.~\ref{sec:results3d} we consider three dimensional systems, where the coupling constant is $g_{d=3}\equiv g_{3D} = 4\pi\hbar^2 a/m$ and $a$ is the scattering length. In the Sec.~\ref{sec:results2d} devoted to two dimensional systems we use notation $g_{d=2} \equiv g_{2D}$. In the case of a box potential with periodic boundary conditions the macroscopically occupied orbital  (the BEC wavefunction in 3D) is just a constant function i.e. a plane wave with momentum $0$. 

In what follows we focus on the fluctuations of the number of atoms in the BEC at finite temperature. We use the canonical and the microcanonical ensembles, which were shown to be close to the experimental reality~\cite{Kristensen2019, Christensen2021}.

There are several different ways to describe the statistical properties of ultra-cold Bose gases theoretically. In this paper, we sample many-body states to generate a set of copies that properly approximates the canonical ensemble of a gas. Given a sufficiently large set of copies, the expectation values are defined as the average over the set. By post-selecting the set we obtain results in the microcanonical ensemble.
\begin{figure}
\fbox{\includegraphics[width=0.5\textwidth]{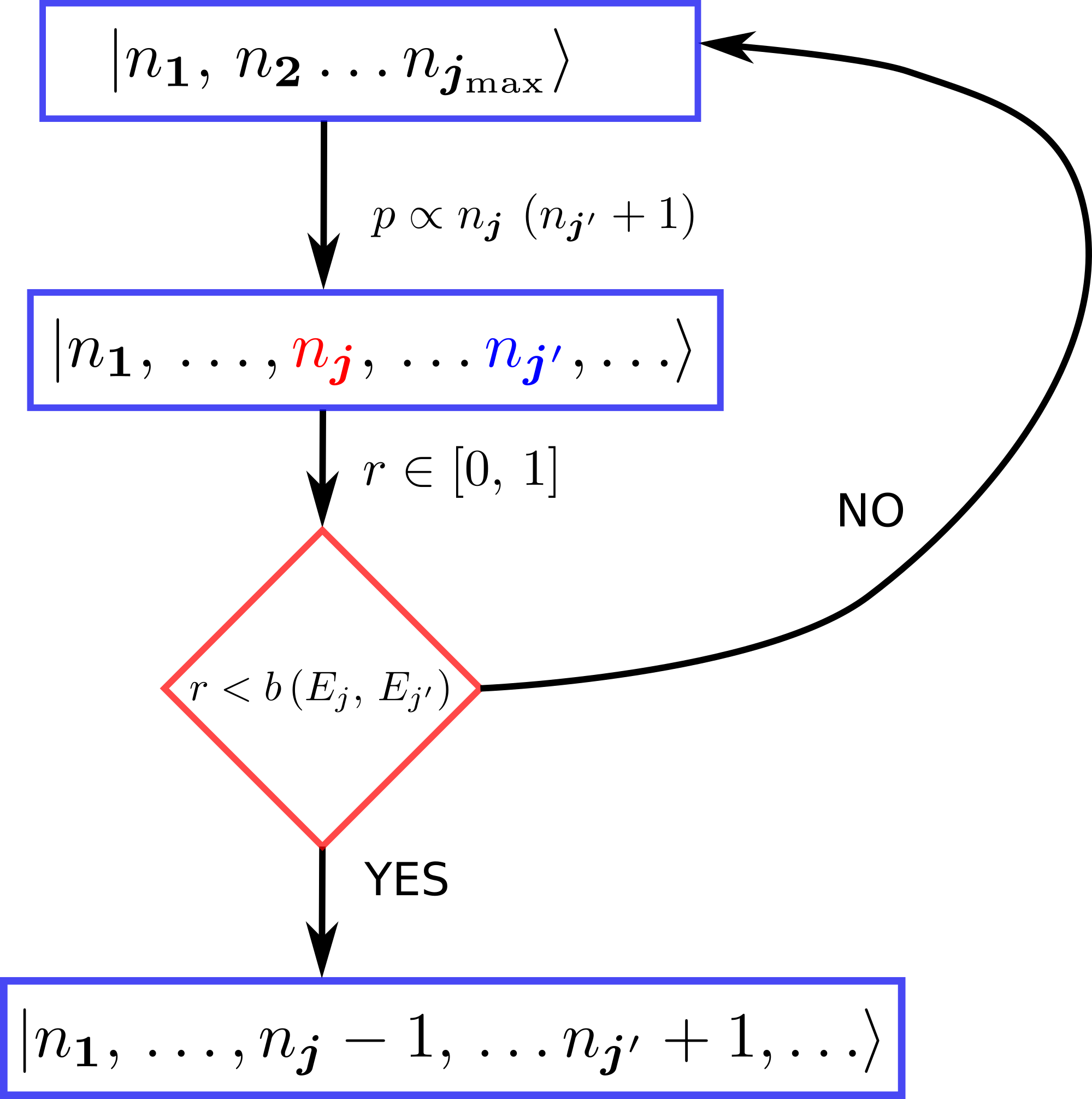}}
\caption{Single step of the FSS method: one draws two states -- one from which an atom might be taken (index $\bm{j}$) and one in which the atom may land (index $\bm{j}^{\prime}$) with probability distribution proportional to $n_{\bm{j}}\bb{n_{\bm{j}^{\prime}}+1}$. The new state is accepted only if a random number $r$ drawn from a uniform distribution in $[0,\,1]$ is smaller than the Boltzman factor $b$ given in Eq.~(\ref{eq:boltzmann-factor}).
\label{fig:FSSM}}
\end{figure}
To define the appropriate Metropolis algorithm~\cite{Metropolis1953} we need to define "the stage", that is the set of available states, and the "Metropolis dynamics" or the specific algorithm defining the Markov chain generating the approximation to the canonical ensemble.

{\it Setting "the stage":} All states of $N$ particles belong to the suitable Hilbert space. A convenient parametrization is provided by the basis of single particle states in the trapping potential. Since we consider box potentials with periodic boundary conditions the basis states are just plane waves
\begin{equation}
    f_{\bm{j}}(\bm{r}) := f_{j_x \,j_y\, j_z}(x,\,y,\,z) = \frac{1}{\sqrt{L_x\, L_y\,L_z}}\,e^{- i\, 2\pi \left(j_x\, x/L_x + j_y\, y /L_y + j_z\, z/L_z\right)},
\end{equation}
where $L_x$, $L_y$, and $L_z$ are the lengths of the box potential and  $j_x$, $j_y$ and $j_z$ are positive and negative integers. It is worth stressing that due to translational invariance, these states remain eigenstates of the single-particle density matrix also for the interacting gas. Thus, the constant function $j_x=j_y=j_z=0$ remains the condensate state also in the presence of interactions.

The space of all $N$-particle states is spanned by the Fock states
\begin{equation}
    \ket{\bm{n}} := \ket{n_{\bm{1}},\,n_{\bm{2}},\,\ldots},
    \label{eq:Fock}
\end{equation}
where $n_{\bm{j}}$ denotes the number of bosons in the single-particle state $f_{\bm{j}}(\bm{r})$. In the canonical ensemble, we fix the total number of atoms $N$ and consider only the Fock states that contain $N$ bosons
\begin{equation}
\sum_{\bm{j}} n_{\bm{j}}= N.
\end{equation}
The whole Hilbert space contains all superpositions of all $N$-particle Fock states. The appropriate parameters are far too numerous for any efficient numerics. Instead, we restrict our set of available states just to the Fock states in Eq.~(\ref{eq:Fock}), not accounting for their superpositions. 
This has two consequences. First, it neglects the phenomenon of quantum depletion. Thus, the method is expected to yield correct results only for weak interactions. Second, it is not applicable to weakly interacting bosons confined in a harmonic trap, since, in this case, the condensate wavefunction is a superposition of many oscillator states.

{\it Metropolis dynamics:} The following algorithm defines our Markov chain used to generate the elements of our representation of the canonical ensemble.  A single step of this algorithm is also shown in Fig.~\ref{fig:FSSM}.

Each particle has the same probability of jumping out of a given single-particle state. The probability of jumping out is proportional to the number of particles in that state. The probability of landing in a given single-particle state is proportional to its occupation (stimulated process) plus one (to account for the spontaneous process). The acceptance criterion, usual for the Metropolis algorithm, is based on comparing a random number $0<r<1$ drawn from a uniform distribution versus the Boltzmann factor $b$ of the initial and the final states
\begin{equation}
b\bb{\ebefore, \eafter} = \exp\bb{-\beta\bb{\ebefore - \eafter}},
\label{eq:boltzmann-factor}
\end{equation}
where $\beta=1/\left(k_BT\right)$, $k_B$ is the Boltzmann constant and $T$ is the temperature. The energy $\ebefore$ is the expectation value of the Hamiltonian in the current Fock state. It is the sum of the kinetic energy and the interaction energy. The kinetic energy is simply
\begin{equation}
    E_{\rm kin} = \sum_{\bm{j}} \,n_{\bm{j}} \,e_{\bm{j}},
\end{equation}
where $e_{\bm{j}}$ is the energy of the $\bm{j}$th level, i.e.
\begin{equation}
    e_{\bm{j}} := \frac{2\pi^2\hbar^2}{m} \left( \bb{\frac{j_x}{L_1}}^2 + \bb{\frac{j_y}{L_2}}^2+\bb{\frac{j_z}{L_3}}^2\right).
\end{equation}
The short-range interaction energy, which in the general case is a nontrivial quadratic form, reduces to a single sum in the case of a box potential and averaged in a single Fock state
\begin{equation}
    E_{\rm int} = \left\langle  {\bm{n}} \left| \frac{g_{3D}}{2}\int\,{\rm d}^3 r\, \oppsis(\bm{r}) \oppsis (\bm{r}) \oppsi(\bm{r}) \oppsi(\bm{r})\right| \bm{n}\right\rangle = \frac{g_{3D}}{2}\left(2 N(N-1) - \sum_{\bm{j}} n_{\bm{j}}^2\right).
\end{equation}
Moreover, for comparison of the Boltzmann factors, only the difference of the energies of the final and initial state enters (see Eq.~(\ref{eq:boltzmann-factor}))
\begin{equation}
    \ebefore - \eafter = e_{\bm{j}} - e_{\bm{j}^{\prime}}
    +g_{3D} \bb{n_{\bm{j}^{\prime}}  - n_{\bm{j}}},
\end{equation}
where $\bm{j}$ and $\bm{j}^{\prime}$ are the indices of the single particle states from which the atom escaped and in which it lands, respectively. A single step of this algorithm is presented in Fig.~\ref{fig:FSSM}.

The algorithm satisfies the detailed balance principle and guarantees access to all important $N$ particle states. 
When the number of steps goes to infinity the expectation value of any physical quantity does not depend on the state used to initiate the algorithm. In practice we perform only a finite number of steps and discard approximately $10\ N$ initials steps during which the quantities of interest are not only fluctuating but also drifting.

Importantly, the Fock State Sampling method also offers access to the microcanonical ensemble. This is simply accomplished by reducing the number of states to those with energy in the small interval around the most probable one.

In the following we present the results obtained with the FSSM for a gas in a 3D (Sec.~\ref{sec:results3d}) and 2D (Sec.~\ref{sec:results2d}) box potential, including interactions between the atoms.

\section{Characteristic temperature for a gas in 3D box potential\label{sec:results3d}}

\begin{figure}
\includegraphics[width=0.8\textwidth]{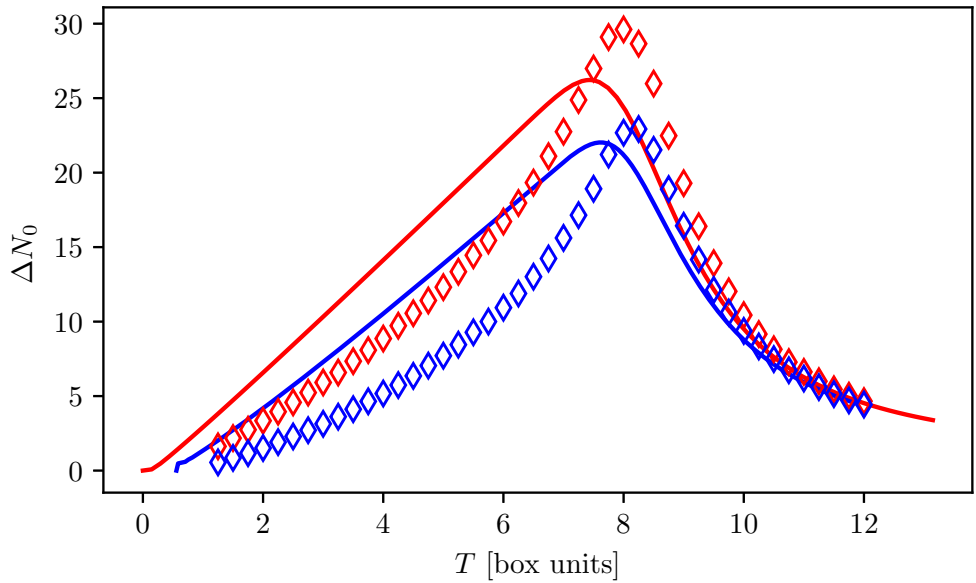}
\caption{Standard deviation of the number of condensed atoms confined in a 3D box potential with periodic boundary conditions for $N=300$ atoms for a non-interacting (solid lines) and an interacting gas with a gas parameter $a \rho^{\frac{1}{3}} = 0.05$ (empty diamonds). Red symbols indicate canonical and blue microcanonical results. The solid lines are exact, obtained with the help of recurrence relations. The empty diamonds represent canonical and microcanonical results obtained with the help of the Fock State Sampling method. Note similar shifts of the characteristic temperature in both statistical ensembles. Also, note that the maximal fluctuations are increased by interactions.
\label{fig:micro}}
\end{figure}

We illustrate the results of our method for ultracold gases of bosonic atoms in a 3D box potential in Fig.~\ref{fig:micro}. The figure shows the temperature dependence of the standard deviation of the number of atoms in a Bose-Einstein condensate $\Delta N_0$ for both canonical and microcanonical ensembles. The results for the non-interacting gas are exact and obtained with the recurrence relations 
while the results for the interacting gas are obtained with the FSS method described in the previous section. Note that the microcaonical ensemble yields significantly lower fluctuations. Moreover, interactions increase the peak fluctuations in both ensembles and similarly the temperature of the maximal fluctuations is increased.

The related shift of the critical temperature due to collisions in weakly interacting Bose gas in a 3D box potential has been the subject of a longstanding debate as outlined in the introduction. The final result for the correction was obtained with a sophisticated numerical method~\cite{Kashurnikov2001}, based on techniques developed over the past 20 years. On the contrary, our method, although approximate, is simple to implement for as many as $10.000$ atoms.

Here, we study the temperature of the maximal fluctuations~$\tchar$ instead of looking at the critical temperature $T_c$. This characteristic temperature is well defined for systems with a finite number of atoms. Moreover, it was recently shown that it can be measured for a Bose gas in a harmonic trap~\cite{Kristensen2019, Christensen2021}. 

Thus, the main quantity of interest is the interaction-induced relative shift to the characteristic temperature $\tchar$
\begin{equation}
\delta \tchar(N,\,a) := \frac{\tchar (N,\,a) - \tchar (N,\,0)}{\tchar (N,\,0)},
\end{equation}
where $\tchar (N,\,a)$ is the characteristic temperature for a gas with $N$ atoms interacting with the $s$-wave scattering length $a$. To find the dependence of $\delta \tchar$ on $N$ and $a$ we study the system with atom numbers ranging from $100$ to $10000$ and interaction strengths $g$ corresponding to gas parameters $a \rho^{\frac{1}{3}}$ from $0.005$ to $0.02$. 
All temperatures are given in units of $2\pi^2\hbar^2/(m k_{\rm B} L^2 )$.
\begin{figure}
\includegraphics{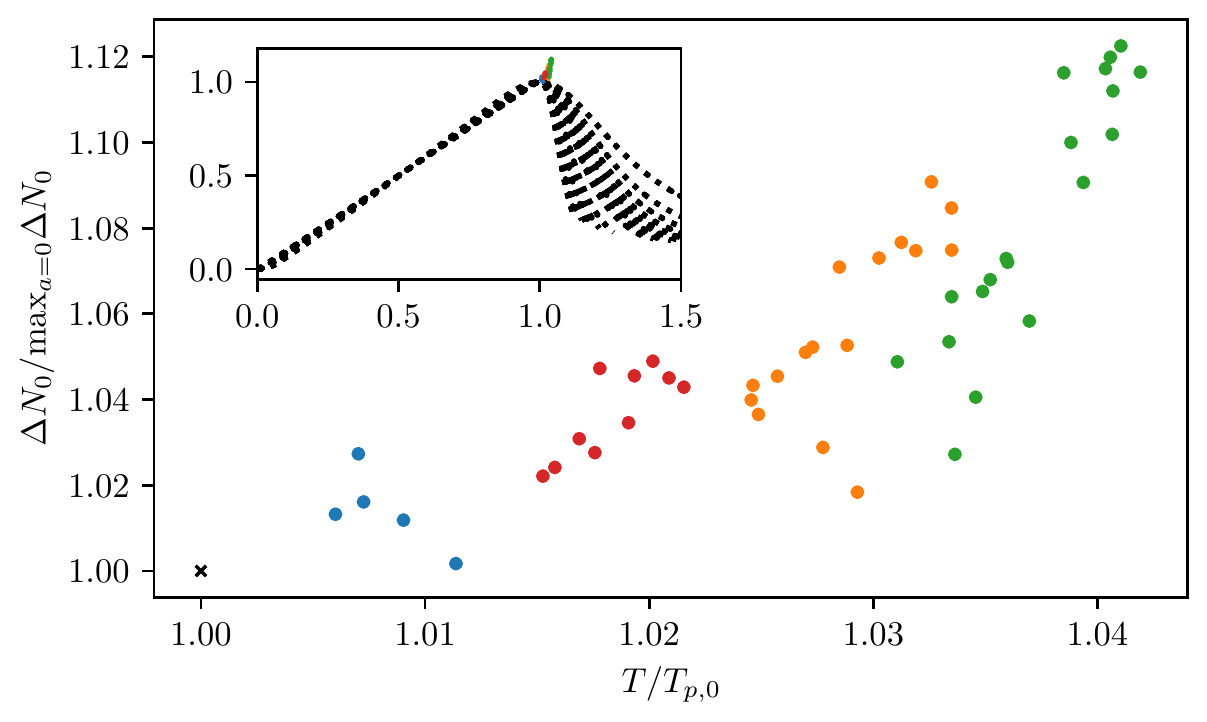}
\caption{Relative standard deviation of the BEC atom number for 
an interacting gas (coloured points) for various total numbers of atoms, and interaction strengths. All points are rescaled by the maximal value of the fluctuations in the non-interacting case. Results for systems with the same gas parameter are marked with the same colour, i.e.  $a \rho^{\frac{1}{3}}$ corresponds to $0.005$ (blue), $0.01$ (red), $0.015$ (orange), $0.02$ (green) . The quantity $T_{{\rm p},0}$ is the temperature of maximal fluctuations of the non-interacting gas and the symbol "X" marks the reference point -- the maximal BEC fluctuations of the non-interacting gas.  The atom number is
in the range  $N=100,200,...,1000,2000,...,10000$.  The inset shows an overview of the entire temperature range with the results for the non-interacting gas (dashed lines).
\label{fig:shift3D}}
\end{figure}

The results are illustrated in Fig.~\ref{fig:shift3D}, which shows the relative standard deviation of the number of condensed atoms as a function of temperature for different total numbers of atoms and various interaction strengths $g$. Since the main focus is the shift of the characteristic temperature $\tchar$, all results for the non-interacting gas are normalized in terms of both, the maximal value and its temperature. The same scaling factors are used for the results for the interacting gas, i.e. the temperature (relative standard deviation) is divided by the temperature $T_{{\rm p}, 0}$ (maximal relative standard deviation) of the non-interacting gas with the same number of atoms. After rescaling one can easily follow the interaction-induced shits. The points with a given color show the maximal variance at the characteristic temperature for various atom numbers $N$ and scattering lengths $a$ but a common gas parameter $a \rho^{\frac{1}{3}}$.

Note that the maximal relative standard deviations are grouped into small regions for a common gas parameter. For larger gas parameters corresponding to larger interactions the maximal relative standard deviations are larger and are reached at higher temperatures as compared to the non-interacting gas.

We fit the average shift for each characteristic temperature with a linear dependence on the gas parameter and obtain 
\begin{equation}
\delta \tchar \approx (2.039\pm0.014) \bb{a \rho^{1/3}},
\end{equation}
in the range of the number of atoms $N$ between $4000$ and $10000$.
Thus the scaling is similar to the one obtained for the critical temperature of an infinitely large system, while the prefactor is larger. 

Note that the maximal relative standard deviations for a common gas parameter form  elongated regions, indicating that the maximal variance and the characteristic temperature may have a further dependence on the scattering length and density. The remaining spread of the points indicates the precision of our method. 

Also note that interactions increase the maximal fluctuations in this case. This point has also been the subject of a long standing controversy (see, for instance, the inset in Fig.~4 in Ref.~\cite{Kristensen2019}). It was recently addressed using the FSS method~\cite{Kruk2022Sep}, showing that the size of the fluctuations depend on all system parameters and thus it is not possible to generalize the effect if interactions on the magnitude condensate fluctuations.

Importantly, the characteristic temperature discussed in this section is also well-defined for systems that do not exhibit a phase transition  in a thermodynamic limit. An example of such a system is a gas in a 2D box potential, discussed in the next section.

\section{Characteristic temperature  in a 2D box potential\label{sec:results2d}}
\begin{figure}
\includegraphics[width=1\textwidth]{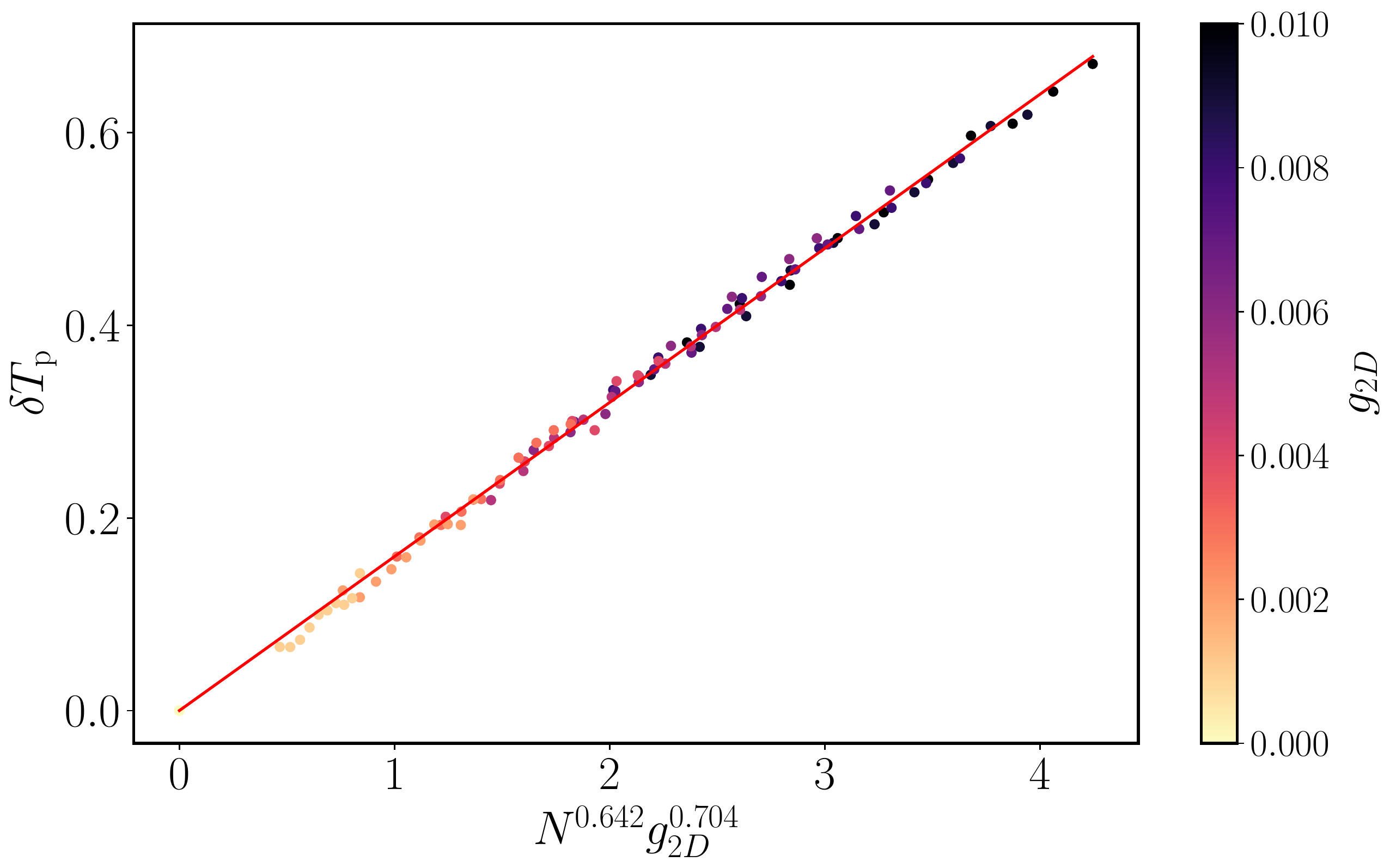}
\caption{Relative shift of the characteristic temperature for Bose gases in a 2D box potential with periodic boundary conditions due to interactions. The coupling constant $g_{2D}$ was varied from $0$ to $0.01$ and the number of atoms was adjusted from $600$ to $1500$.
\label{fig:fluct2d}}
\end{figure}
It is well known, that Bose-Einstein condensation appears as a phase transition for sufficiently high dimensions. 
In a box potential the phase transition occurs only in three dimensions, while it is absent in lower dimensionality. Despite the fact that there is no phase transition and therefore no critical temperature in the case of two dimensions, the notion of the characteristic temperature $\tchar$, marking the temperature of maximal fluctuations, is still applicable. Of course in the absence of the phase transition, interactions still affect the fluctuations. 

We illustrate this by investigating a two-dimensional box potential with periodic boundary conditions. The calculation is analogous to the 3D box potential and the condensate wave function is still the constant one, regardless of interaction. The algorithm, after omitting all $Z$ dependent variables, is identical to the one introduced in the previous section. The relative shift of the characteristic temperature due to interactions was calculated for various atom numbers $N$ and interaction strengths $g_{2D}$ as defined in the Hamiltonian Eq.~(\ref{eq:hamiltonian}). In this Section we do not refer to the gas parameter, which would be more complicated than in the 3D case.

Figure~\ref{fig:fluct2d} shows this shift as a function of $g_{2D}^\alpha N^\beta$ with optimally chosen exponents $\alpha$ and $\beta$ obtained from a fit to the data.  yielding
\begin{equation}
\delta \tchar^{(2D)}(N,\,g_{2D}) := \bb{0.16\pm0.03} N^{0.642\pm0.015} g_{2D}^{0.704\pm0.006}.
\end{equation}
The stated errors may be reduced at the expense of the numerical effort. The convergence is very slow and the errors scale with the square root of the number of Metropolis steps. 

The results presented in this section illustrate of the power and generality of the FSS method. The method is conceptually very simple and its successful application merely requires a numerical effort.

\section{Conclusion and outlook\label{sec:conclusions}}

In conclusion, we have investigated the fluctuations of ideal and weakly interacting the Bose-Einstein condensates trapped in box potentials with periodic boundary conditions. The temperature of maximal BEC atom number fluctuations $\tchar$ was analysed under various conditions.  The advantage of $\tchar$ over $T_c$ lies in the fact, that it is unambiguously defined also for a finite system, and it may be studied also in systems that do not exhibit phase transition. 

In our study, we used the Fock State Sampling method, which turns out to be easy to use, exact for the non-interacting system (see~\cite{Kruk2022Sep}) and applicable for wide range of problems. With this method, we found the shift of the characteristic temperature in the 3D box potential to be $\approx 2.03\,\left(a\,\rho^{1/3}\right)$, where $a$ is the scattering length and $\rho$ is a gas density. This is reasonably close to the expected shift of the critical temperature in this system $\approx 1.3\,\left(a\,\rho^{1/3}\right)$.

We also applied our method to a two-dimensional system and obtained a generalized dependence of the characteristic temperature on the interaction strength and atom number $\approx 0.16 N^{0.642} g_{2D}^{0.704}$, showing the applicability in a system that does not exhibit a phase transition.

Experimentally, the recent realization of box potentials provides an opportunity to address the predictions presented above. In particular, combining box potentials with atomic species that allow for tunability of the interaction strength will provide access to a wide variation of gas parameters.

Box potentials are typically created using blue-detuned light to form the walls of a box, such as e.g. a hollow beam with two narrow light sheets as end caps~\cite{navon_quantum_2021}. To flatten the bottom of the potential, gravity must be compensated using a magnetic field gradient. Alternatively, a light field with a linearly varying intensity produced by an accousto-optic deflector can be used~\cite{Shibata2020}. The necessary beam shapes for box potentials can be generated using spatial light modulators, digital micromirror devices or specialized optical elements such as axicons. 

Tunability of the scattering length would be highly beneficial to isolate the effect of interactions on the characteristic temperature and the magnitude of atom number fluctuations. This can be achieved by adjusting the magnetic field near a Feshbach resonance~\cite{chin_feshbach_2010}. Since a magnetic field gradient is the most common way to cancel gravity within a box potential this will typically necessitate independent control of the magnetic field gradient and its mean value. The field gradient will thus introduce a spatial dependence of the scattering length and hence atomic species with broad resonances such as e.g. the bosonic isotopes of potassium should be used.


Furthermore, it is important to distinguish between the BEC and the thermal part of partially condensed atomic clouds to measure the atom number fluctuations. Thus, the bimodality of the momentum distribution is crucial for determining the BEC number and the number of thermal atoms. Fortunately, both the bimodality and an appropriate fitting function for the thermal cloud have been confirmed~\cite{Gaunt2013} experimentally.

The most significant outstanding challenge towards the measurement of fluctuations proposed here is the combination of box potentials with atom number stabilisation. There are two primary technical sources of variations in the total number of atoms. The first one is due to the statistical nature of evaporative cooling, which relates the atom number to the temperature. This is predictable and can be accounted for in the evaluation of atom number fluctuations. The second source of atom number variation is typically due to various technical noise sources in the experiment and should be minimized since it can distort the measured atom number fluctuation when different mean values of the BEC atom number are probed. 
Thus, to conduct the experiment, it will be necessary to combine the box potential with atom number stabilization. However, it is not yet clear whether it is sufficient to stabilize the atom number before loading the cloud into the box potential, or if methods for stabilization within the box potential must be developed.

The combination of the Fock State Sampling method and current experimental developments will allow for further experiments in the near future. Especially,  since the FSSM can provide precise predictions for experimentally relevant atom numbers in a variety of potentials, the time has now come for a new generation of experiments on these fundamental questions.

\section*{Acknowledgements}

We dedicate this paper to Professor Iwo Białynicki-Birula on the occasion of his 90$th$ birthday. The Polish authors represent three generations of his students. We owe him a lot in terms of knowledge and style of doing research. Moreover, we pursue our research careers at the Center for Theoretical Physics PAS founded by IBB. 
Working at the institute, under the influence of IBB, formed us as physicists. His papers are exceptionally clearly written and we strive to be similarly clear in our work.

We thank P. Deuar for fruitful discussions.

\paragraph{Funding information}
M.~B.~K. acknowledges support from the (Polish) National Science Center Grants No.~2018/31/B/ST2/01871 and No.~2022/45/N/ST2/03511.
K.~P. acknowledges support from the (Polish) National Science Center Grant No.~2019/34/E/ST2/00289. K.~R., P.~K. and M.~B.~K. acknowledge support from the (Polish) National Science Center Grant No.~2021/43/B/ST2/01426. 
Center for Theoretical Physics of the Polish Academy of Sciences is a member of the National Laboratory of Atomic, Molecular and Optical Physics (KL FAMO).

T.~V. and J.~A. acknowledge support from the Danish National Research Foundation through the Center of Excellence (Grant Agreement No. DNRF156) and from the Independent Research Fund Denmark-Natural Sciences via Grant No.~8021-00233B and 0135-00205B.

\bibliography{bibliography.bib}

\begin{thebibliography}{24}%
\makeatletter
\providecommand \@ifxundefined [1]{%
 \@ifx{#1\undefined}
}%
\providecommand \@ifnum [1]{%
 \ifnum #1\expandafter \@firstoftwo
 \else \expandafter \@secondoftwo
 \fi
}%
\providecommand \@ifx [1]{%
 \ifx #1\expandafter \@firstoftwo
 \else \expandafter \@secondoftwo
 \fi
}%
\providecommand \natexlab [1]{#1}%
\providecommand \enquote  [1]{``#1''}%
\providecommand \bibnamefont  [1]{#1}%
\providecommand \bibfnamefont [1]{#1}%
\providecommand \citenamefont [1]{#1}%
\providecommand \href@noop [0]{\@secondoftwo}%
\providecommand \href [0]{\begingroup \@sanitize@url \@href}%
\providecommand \@href[1]{\@@startlink{#1}\@@href}%
\providecommand \@@href[1]{\endgroup#1\@@endlink}%
\providecommand \@sanitize@url [0]{\catcode `\\12\catcode `\$12\catcode
  `\&12\catcode `\#12\catcode `\^12\catcode `\_12\catcode `\%12\relax}%
\providecommand \@@startlink[1]{}%
\providecommand \@@endlink[0]{}%
\providecommand \url  [0]{\begingroup\@sanitize@url \@url }%
\providecommand \@url [1]{\endgroup\@href {#1}{\urlprefix }}%
\providecommand \urlprefix  [0]{URL }%
\providecommand \Eprint [0]{\href }%
\providecommand \doibase [0]{http://dx.doi.org/}%
\providecommand \selectlanguage [0]{\@gobble}%
\providecommand \bibinfo  [0]{\@secondoftwo}%
\providecommand \bibfield  [0]{\@secondoftwo}%
\providecommand \translation [1]{[#1]}%
\providecommand \BibitemOpen [0]{}%
\providecommand \bibitemStop [0]{}%
\providecommand \bibitemNoStop [0]{.\EOS\space}%
\providecommand \EOS [0]{\spacefactor3000\relax}%
\providecommand \BibitemShut  [1]{\csname bibitem#1\endcsname}%
\let\auto@bib@innerbib\@empty
\bibitem [{\citenamefont {Lee}\ and\ \citenamefont {Yang}(1957)}]{Lee1957}%
  \BibitemOpen
  \bibfield  {author} {\bibinfo {author} {\bibfnamefont {T.~D.}\ \bibnamefont
  {Lee}}\ and\ \bibinfo {author} {\bibfnamefont {C.~N.}\ \bibnamefont {Yang}},\
  }\href {\doibase 10.1103/PhysRev.105.1119} {\bibfield  {journal} {\bibinfo
  {journal} {Phys. Rev.}\ }\textbf {\bibinfo {volume} {105}},\ \bibinfo {pages}
  {1119} (\bibinfo {year} {1957})}\BibitemShut {NoStop}%
\bibitem [{\citenamefont {Gr\"uter}\ \emph {et~al.}(1997)\citenamefont
  {Gr\"uter}, \citenamefont {Ceperley},\ and\ \citenamefont
  {Lalo\"e}}]{Grueter1997}%
  \BibitemOpen
  \bibfield  {author} {\bibinfo {author} {\bibfnamefont {P.}~\bibnamefont
  {Gr\"uter}}, \bibinfo {author} {\bibfnamefont {D.}~\bibnamefont {Ceperley}},
  \ and\ \bibinfo {author} {\bibfnamefont {F.}~\bibnamefont {Lalo\"e}},\ }\href
  {\doibase 10.1103/PhysRevLett.79.3549} {\bibfield  {journal} {\bibinfo
  {journal} {Phys. Rev. Lett.}\ }\textbf {\bibinfo {volume} {79}},\ \bibinfo
  {pages} {3549} (\bibinfo {year} {1997})}\BibitemShut {NoStop}%
\bibitem [{\citenamefont {Holzmann}\ \emph {et~al.}(1999)\citenamefont
  {Holzmann}, \citenamefont {Gr{\ifmmode\ddot{u}\else\"{u}\fi}ter},\ and\
  \citenamefont {Lalo{\ifmmode\ddot{e}\else\"{e}\fi}}}]{Holzmann1999Aug}%
  \BibitemOpen
  \bibfield  {author} {\bibinfo {author} {\bibfnamefont {M.}~\bibnamefont
  {Holzmann}}, \bibinfo {author} {\bibfnamefont {P.}~\bibnamefont
  {Gr{\ifmmode\ddot{u}\else\"{u}\fi}ter}}, \ and\ \bibinfo {author}
  {\bibfnamefont {F.}~\bibnamefont {Lalo{\ifmmode\ddot{e}\else\"{e}\fi}}},\
  }\href {\doibase 10.1007/s100510050905} {\bibfield  {journal} {\bibinfo
  {journal} {Eur. Phys. J. B}\ }\textbf {\bibinfo {volume} {10}},\ \bibinfo
  {pages} {739} (\bibinfo {year} {1999})}\BibitemShut {NoStop}%
\bibitem [{\citenamefont {Holzmann}\ and\ \citenamefont
  {Krauth}(1999)}]{holzman1999}%
  \BibitemOpen
  \bibfield  {author} {\bibinfo {author} {\bibfnamefont {M.}~\bibnamefont
  {Holzmann}}\ and\ \bibinfo {author} {\bibfnamefont {W.}~\bibnamefont
  {Krauth}},\ }\href {\doibase 10.1103/PhysRevLett.83.2687} {\bibfield
  {journal} {\bibinfo  {journal} {Phys. Rev. Lett.}\ }\textbf {\bibinfo
  {volume} {83}},\ \bibinfo {pages} {2687} (\bibinfo {year}
  {1999})}\BibitemShut {NoStop}%
\bibitem [{\citenamefont {Baym}\ \emph {et~al.}(1999)\citenamefont {Baym},
  \citenamefont {Blaizot}, \citenamefont {Holzmann}, \citenamefont {Lalo\"e},\
  and\ \citenamefont {Vautherin}}]{Baym1999Aug}%
  \BibitemOpen
  \bibfield  {author} {\bibinfo {author} {\bibfnamefont {G.}~\bibnamefont
  {Baym}}, \bibinfo {author} {\bibfnamefont {J.-P.}\ \bibnamefont {Blaizot}},
  \bibinfo {author} {\bibfnamefont {M.}~\bibnamefont {Holzmann}}, \bibinfo
  {author} {\bibfnamefont {F.}~\bibnamefont {Lalo\"e}}, \ and\ \bibinfo
  {author} {\bibfnamefont {D.}~\bibnamefont {Vautherin}},\ }\href {\doibase
  10.1103/PhysRevLett.83.1703} {\bibfield  {journal} {\bibinfo  {journal}
  {Phys. Rev. Lett.}\ }\textbf {\bibinfo {volume} {83}},\ \bibinfo {pages}
  {1703} (\bibinfo {year} {1999})}\BibitemShut {NoStop}%
\bibitem [{\citenamefont {Wilkens}\ \emph {et~al.}(2000)\citenamefont
  {Wilkens}, \citenamefont {Illuminati},\ and\ \citenamefont
  {Kr{\ifmmode\ddot{a}\else\"{a}\fi}mer}}]{Wilkens2000Oct}%
  \BibitemOpen
  \bibfield  {author} {\bibinfo {author} {\bibfnamefont {M.}~\bibnamefont
  {Wilkens}}, \bibinfo {author} {\bibfnamefont {F.}~\bibnamefont {Illuminati}},
  \ and\ \bibinfo {author} {\bibfnamefont {M.}~\bibnamefont
  {Kr{\ifmmode\ddot{a}\else\"{a}\fi}mer}},\ }\href {\doibase
  10.1088/0953-4075/33/20/10j} {\bibfield  {journal} {\bibinfo  {journal} {J.
  Phys. B: At. Mol. Opt. Phys.}\ }\textbf {\bibinfo {volume} {33}},\ \bibinfo
  {pages} {L779} (\bibinfo {year} {2000})}\BibitemShut {NoStop}%
\bibitem [{\citenamefont {Arnold}\ and\ \citenamefont
  {Tom\'a\ifmmode~\check{s}\else \v{s}\fi{}ik}(2000)}]{Arnold2000}%
  \BibitemOpen
  \bibfield  {author} {\bibinfo {author} {\bibfnamefont {P.}~\bibnamefont
  {Arnold}}\ and\ \bibinfo {author} {\bibfnamefont {B.}~\bibnamefont
  {Tom\'a\ifmmode~\check{s}\else \v{s}\fi{}ik}},\ }\href {\doibase
  10.1103/PhysRevA.62.063604} {\bibfield  {journal} {\bibinfo  {journal} {Phys.
  Rev. A}\ }\textbf {\bibinfo {volume} {62}},\ \bibinfo {pages} {063604}
  (\bibinfo {year} {2000})}\BibitemShut {NoStop}%
\bibitem [{\citenamefont {Baym}\ \emph {et~al.}(2000)\citenamefont {Baym},
  \citenamefont {Blaizot},\ and\ \citenamefont {Zinn-Justin}}]{Baym2000Jan}%
  \BibitemOpen
  \bibfield  {author} {\bibinfo {author} {\bibfnamefont {G.}~\bibnamefont
  {Baym}}, \bibinfo {author} {\bibfnamefont {J.-P.}\ \bibnamefont {Blaizot}}, \
  and\ \bibinfo {author} {\bibfnamefont {J.}~\bibnamefont {Zinn-Justin}},\
  }\href {\doibase 10.1209/epl/i2000-00130-3} {\bibfield  {journal} {\bibinfo
  {journal} {Europhys. Lett.}\ }\textbf {\bibinfo {volume} {49}},\ \bibinfo
  {pages} {150} (\bibinfo {year} {2000})}\BibitemShut {NoStop}%
\bibitem [{\citenamefont {de~Souza~Cruz}\ \emph {et~al.}(2001)\citenamefont
  {de~Souza~Cruz}, \citenamefont {Pinto},\ and\ \citenamefont
  {Ramos}}]{Cruz2001}%
  \BibitemOpen
  \bibfield  {author} {\bibinfo {author} {\bibfnamefont {F.~F.}\ \bibnamefont
  {de~Souza~Cruz}}, \bibinfo {author} {\bibfnamefont {M.~B.}\ \bibnamefont
  {Pinto}}, \ and\ \bibinfo {author} {\bibfnamefont {R.~O.}\ \bibnamefont
  {Ramos}},\ }\href {\doibase 10.1103/PhysRevB.64.014515} {\bibfield  {journal}
  {\bibinfo  {journal} {Phys. Rev. B}\ }\textbf {\bibinfo {volume} {64}},\
  \bibinfo {pages} {014515} (\bibinfo {year} {2001})}\BibitemShut {NoStop}%
\bibitem [{\citenamefont {Kashurnikov}\ \emph {et~al.}(2001)\citenamefont
  {Kashurnikov}, \citenamefont {Prokof'ev},\ and\ \citenamefont
  {Svistunov}}]{Kashurnikov2001}%
  \BibitemOpen
  \bibfield  {author} {\bibinfo {author} {\bibfnamefont {V.~A.}\ \bibnamefont
  {Kashurnikov}}, \bibinfo {author} {\bibfnamefont {N.~V.}\ \bibnamefont
  {Prokof'ev}}, \ and\ \bibinfo {author} {\bibfnamefont {B.~V.}\ \bibnamefont
  {Svistunov}},\ }\href {\doibase 10.1103/PhysRevLett.87.120402} {\bibfield
  {journal} {\bibinfo  {journal} {Phys. Rev. Lett.}\ }\textbf {\bibinfo
  {volume} {87}},\ \bibinfo {pages} {120402} (\bibinfo {year}
  {2001})}\BibitemShut {NoStop}%
\bibitem [{\citenamefont {Andersen}(2004)}]{Andersen2004}%
  \BibitemOpen
  \bibfield  {author} {\bibinfo {author} {\bibfnamefont {J.~O.}\ \bibnamefont
  {Andersen}},\ }\href {\doibase 10.1103/RevModPhys.76.599} {\bibfield
  {journal} {\bibinfo  {journal} {Rev. Mod. Phys.}\ }\textbf {\bibinfo {volume}
  {76}},\ \bibinfo {pages} {599} (\bibinfo {year} {2004})}\BibitemShut
  {NoStop}%
\bibitem [{\citenamefont {Davis}\ and\ \citenamefont
  {Morgan}(2003)}]{Davis2003Nov}%
  \BibitemOpen
  \bibfield  {author} {\bibinfo {author} {\bibfnamefont {M.~J.}\ \bibnamefont
  {Davis}}\ and\ \bibinfo {author} {\bibfnamefont {S.~A.}\ \bibnamefont
  {Morgan}},\ }\href {\doibase 10.1103/PhysRevA.68.053615} {\bibfield
  {journal} {\bibinfo  {journal} {Phys. Rev. A}\ }\textbf {\bibinfo {volume}
  {68}},\ \bibinfo {pages} {053615} (\bibinfo {year} {2003})}\BibitemShut
  {NoStop}%
\bibitem [{\citenamefont {Nho}\ and\ \citenamefont
  {Landau}(2004)}]{kwangsik2004}%
  \BibitemOpen
  \bibfield  {author} {\bibinfo {author} {\bibfnamefont {K.}~\bibnamefont
  {Nho}}\ and\ \bibinfo {author} {\bibfnamefont {D.~P.}\ \bibnamefont
  {Landau}},\ }\href {\doibase 10.1103/PhysRevA.70.053614} {\bibfield
  {journal} {\bibinfo  {journal} {Phys. Rev. A}\ }\textbf {\bibinfo {volume}
  {70}},\ \bibinfo {pages} {053614} (\bibinfo {year} {2004})}\BibitemShut
  {NoStop}%
\bibitem [{\citenamefont {Watabe}\ and\ \citenamefont
  {Ohashi}(2013)}]{Watabe2013}%
  \BibitemOpen
  \bibfield  {author} {\bibinfo {author} {\bibfnamefont {S.}~\bibnamefont
  {Watabe}}\ and\ \bibinfo {author} {\bibfnamefont {Y.}~\bibnamefont
  {Ohashi}},\ }\href {\doibase 10.1103/PhysRevA.88.053633} {\bibfield
  {journal} {\bibinfo  {journal} {Phys. Rev. A}\ }\textbf {\bibinfo {volume}
  {88}},\ \bibinfo {pages} {053633} (\bibinfo {year} {2013})}\BibitemShut
  {NoStop}%
\bibitem [{\citenamefont {Kruk}\ \emph {et~al.}(2022)\citenamefont {Kruk},
  \citenamefont {Hryniuk}, \citenamefont {Kristensen}, \citenamefont {Vibel},
  \citenamefont {Paw{\l}owski}, \citenamefont {Arlt},\ and\ \citenamefont
  {Rz{\k{a}}{\ifmmode\dot{z}\else\.{z}\fi}ewski}}]{Kruk2022Sep}%
  \BibitemOpen
  \bibfield  {author} {\bibinfo {author} {\bibfnamefont {M.~B.}\ \bibnamefont
  {Kruk}}, \bibinfo {author} {\bibfnamefont {D.}~\bibnamefont {Hryniuk}},
  \bibinfo {author} {\bibfnamefont {M.}~\bibnamefont {Kristensen}}, \bibinfo
  {author} {\bibfnamefont {T.}~\bibnamefont {Vibel}}, \bibinfo {author}
  {\bibfnamefont {K.}~\bibnamefont {Paw{\l}owski}}, \bibinfo {author}
  {\bibfnamefont {J.}~\bibnamefont {Arlt}}, \ and\ \bibinfo {author}
  {\bibfnamefont {K.}~\bibnamefont
  {Rz{\k{a}}{\ifmmode\dot{z}\else\.{z}\fi}ewski}},\ }\href
  {https://scipost.org/submissions/scipost_202206_00011v2} {\enquote {\bibinfo
  {title} {{Microcanonical and Canonical Fluctuations in atomic Bose-Einstein
  Condensates -- Fock state sampling approach}},}\ } (\bibinfo {year} {2022}),\
  \bibinfo {note} {[Online; accessed 19. Oct. 2022]}\BibitemShut {NoStop}%
\bibitem [{\citenamefont {Gaunt}\ \emph {et~al.}(2013)\citenamefont {Gaunt},
  \citenamefont {Schmidutz}, \citenamefont {Gotlibovych}, \citenamefont
  {Smith},\ and\ \citenamefont {Hadzibabic}}]{Gaunt2013}%
  \BibitemOpen
  \bibfield  {author} {\bibinfo {author} {\bibfnamefont {A.~L.}\ \bibnamefont
  {Gaunt}}, \bibinfo {author} {\bibfnamefont {T.~F.}\ \bibnamefont
  {Schmidutz}}, \bibinfo {author} {\bibfnamefont {I.}~\bibnamefont
  {Gotlibovych}}, \bibinfo {author} {\bibfnamefont {R.~P.}\ \bibnamefont
  {Smith}}, \ and\ \bibinfo {author} {\bibfnamefont {Z.}~\bibnamefont
  {Hadzibabic}},\ }\href {\doibase 10.1103/PhysRevLett.110.200406} {\bibfield
  {journal} {\bibinfo  {journal} {Phys. Rev. Lett.}\ }\textbf {\bibinfo
  {volume} {110}},\ \bibinfo {pages} {200406} (\bibinfo {year}
  {2013})}\BibitemShut {NoStop}%
\bibitem [{\citenamefont {Idziaszek}\ and\ \citenamefont {Rza\ifmmode
  \mbox{\c{}}\else \c{}\fi{}\ifmmode~\dot{z}\else
  \.{z}\fi{}ewski}(2003)}]{Idziaszek2003}%
  \BibitemOpen
  \bibfield  {author} {\bibinfo {author} {\bibfnamefont {Z.}~\bibnamefont
  {Idziaszek}}\ and\ \bibinfo {author} {\bibfnamefont {K.}~\bibnamefont
  {Rza\ifmmode \mbox{\c{}}\else \c{}\fi{}\ifmmode~\dot{z}\else
  \.{z}\fi{}ewski}},\ }\href {\doibase 10.1103/PhysRevA.68.035604} {\bibfield
  {journal} {\bibinfo  {journal} {Phys. Rev. A}\ }\textbf {\bibinfo {volume}
  {68}},\ \bibinfo {pages} {035604} (\bibinfo {year} {2003})}\BibitemShut
  {NoStop}%
\bibitem [{\citenamefont {Gajdacz}\ \emph {et~al.}(2016)\citenamefont
  {Gajdacz}, \citenamefont {Hilliard}, \citenamefont {Kristensen},
  \citenamefont {Pedersen}, \citenamefont {Klempt}, \citenamefont {Arlt},\ and\
  \citenamefont {Sherson}}]{Gajdacz2016}%
  \BibitemOpen
  \bibfield  {author} {\bibinfo {author} {\bibfnamefont {M.}~\bibnamefont
  {Gajdacz}}, \bibinfo {author} {\bibfnamefont {A.~J.}\ \bibnamefont
  {Hilliard}}, \bibinfo {author} {\bibfnamefont {M.~A.}\ \bibnamefont
  {Kristensen}}, \bibinfo {author} {\bibfnamefont {P.~L.}\ \bibnamefont
  {Pedersen}}, \bibinfo {author} {\bibfnamefont {C.}~\bibnamefont {Klempt}},
  \bibinfo {author} {\bibfnamefont {J.~J.}\ \bibnamefont {Arlt}}, \ and\
  \bibinfo {author} {\bibfnamefont {J.~F.}\ \bibnamefont {Sherson}},\ }\href
  {\doibase 10.1103/PhysRevLett.117.073604} {\bibfield  {journal} {\bibinfo
  {journal} {Phys. Rev. Lett.}\ }\textbf {\bibinfo {volume} {117}},\ \bibinfo
  {pages} {073604} (\bibinfo {year} {2016})}\BibitemShut {NoStop}%
\bibitem [{\citenamefont {Kristensen}\ \emph {et~al.}(2019)\citenamefont
  {Kristensen}, \citenamefont {Christensen}, \citenamefont {Gajdacz},
  \citenamefont {Iglicki}, \citenamefont {Paw{\l}owski}, \citenamefont
  {Klempt}, \citenamefont {Sherson}, \citenamefont {Rza{\.{z}}ewski},
  \citenamefont {Hilliard},\ and\ \citenamefont {Arlt}}]{Kristensen2019}%
  \BibitemOpen
  \bibfield  {author} {\bibinfo {author} {\bibfnamefont {M.}~\bibnamefont
  {Kristensen}}, \bibinfo {author} {\bibfnamefont {M.}~\bibnamefont
  {Christensen}}, \bibinfo {author} {\bibfnamefont {M.}~\bibnamefont
  {Gajdacz}}, \bibinfo {author} {\bibfnamefont {M.}~\bibnamefont {Iglicki}},
  \bibinfo {author} {\bibfnamefont {K.}~\bibnamefont {Paw{\l}owski}}, \bibinfo
  {author} {\bibfnamefont {C.}~\bibnamefont {Klempt}}, \bibinfo {author}
  {\bibfnamefont {J.}~\bibnamefont {Sherson}}, \bibinfo {author} {\bibfnamefont
  {K.}~\bibnamefont {Rza{\.{z}}ewski}}, \bibinfo {author} {\bibfnamefont
  {A.}~\bibnamefont {Hilliard}}, \ and\ \bibinfo {author} {\bibfnamefont
  {J.}~\bibnamefont {Arlt}},\ }\href {\doibase 10.1103/physrevlett.122.163601}
  {\bibfield  {journal} {\bibinfo  {journal} {Phys. Rev. Lett.}\ }\textbf
  {\bibinfo {volume} {122}},\ \bibinfo {pages} {163601} (\bibinfo {year}
  {2019})}\BibitemShut {NoStop}%
\bibitem [{\citenamefont {Christensen}\ \emph {et~al.}(2021)\citenamefont
  {Christensen}, \citenamefont {Vibel}, \citenamefont {Hilliard}, \citenamefont
  {Kruk}, \citenamefont {Paw{\l}owski}, \citenamefont {Hryniuk}, \citenamefont
  {Rza{\.{z}}ewski}, \citenamefont {Kristensen},\ and\ \citenamefont
  {Arlt}}]{Christensen2021}%
  \BibitemOpen
  \bibfield  {author} {\bibinfo {author} {\bibfnamefont {M.}~\bibnamefont
  {Christensen}}, \bibinfo {author} {\bibfnamefont {T.}~\bibnamefont {Vibel}},
  \bibinfo {author} {\bibfnamefont {A.}~\bibnamefont {Hilliard}}, \bibinfo
  {author} {\bibfnamefont {M.}~\bibnamefont {Kruk}}, \bibinfo {author}
  {\bibfnamefont {K.}~\bibnamefont {Paw{\l}owski}}, \bibinfo {author}
  {\bibfnamefont {D.}~\bibnamefont {Hryniuk}}, \bibinfo {author} {\bibfnamefont
  {K.}~\bibnamefont {Rza{\.{z}}ewski}}, \bibinfo {author} {\bibfnamefont
  {M.}~\bibnamefont {Kristensen}}, \ and\ \bibinfo {author} {\bibfnamefont
  {J.}~\bibnamefont {Arlt}},\ }\href {\doibase 10.1103/physrevlett.126.153601}
  {\bibfield  {journal} {\bibinfo  {journal} {Phys. Rev. Lett.}\ }\textbf
  {\bibinfo {volume} {126}},\ \bibinfo {pages} {153601} (\bibinfo {year}
  {2021})}\BibitemShut {NoStop}%
\bibitem [{\citenamefont {Metropolis}\ \emph {et~al.}(1953)\citenamefont
  {Metropolis}, \citenamefont {Rosenbluth}, \citenamefont {Rosenbluth},
  \citenamefont {Teller},\ and\ \citenamefont {Teller}}]{Metropolis1953}%
  \BibitemOpen
  \bibfield  {author} {\bibinfo {author} {\bibfnamefont {N.}~\bibnamefont
  {Metropolis}}, \bibinfo {author} {\bibfnamefont {A.~W.}\ \bibnamefont
  {Rosenbluth}}, \bibinfo {author} {\bibfnamefont {M.~N.}\ \bibnamefont
  {Rosenbluth}}, \bibinfo {author} {\bibfnamefont {A.~H.}\ \bibnamefont
  {Teller}}, \ and\ \bibinfo {author} {\bibfnamefont {E.}~\bibnamefont
  {Teller}},\ }\href {\doibase 10.1063/1.1699114} {\bibfield  {journal}
  {\bibinfo  {journal} {The Journal of Chemical Physics}\ }\textbf {\bibinfo
  {volume} {21}},\ \bibinfo {pages} {1087} (\bibinfo {year}
  {1953})}\BibitemShut {NoStop}%
\bibitem [{\citenamefont {Navon}\ \emph {et~al.}(2021)\citenamefont {Navon},
  \citenamefont {Smith},\ and\ \citenamefont
  {Hadzibabic}}]{navon_quantum_2021}%
  \BibitemOpen
  \bibfield  {author} {\bibinfo {author} {\bibfnamefont {N.}~\bibnamefont
  {Navon}}, \bibinfo {author} {\bibfnamefont {R.~P.}\ \bibnamefont {Smith}}, \
  and\ \bibinfo {author} {\bibfnamefont {Z.}~\bibnamefont {Hadzibabic}},\
  }\href {\doibase 10.1038/s41567-021-01403-z} {\bibfield  {journal} {\bibinfo
  {journal} {Nature Physics}\ }\textbf {\bibinfo {volume} {17}},\ \bibinfo
  {pages} {1334} (\bibinfo {year} {2021})},\ \bibinfo {note} {number: 12
  Publisher: Nature Publishing Group}\BibitemShut {NoStop}%
\bibitem [{\citenamefont {Shibata}\ \emph {et~al.}(2020)\citenamefont
  {Shibata}, \citenamefont {Ikeda}, \citenamefont {Suzuki},\ and\ \citenamefont
  {Hirano}}]{Shibata2020}%
  \BibitemOpen
  \bibfield  {author} {\bibinfo {author} {\bibfnamefont {K.}~\bibnamefont
  {Shibata}}, \bibinfo {author} {\bibfnamefont {H.}~\bibnamefont {Ikeda}},
  \bibinfo {author} {\bibfnamefont {R.}~\bibnamefont {Suzuki}}, \ and\ \bibinfo
  {author} {\bibfnamefont {T.}~\bibnamefont {Hirano}},\ }\href {\doibase
  10.1103/PhysRevResearch.2.013068} {\bibfield  {journal} {\bibinfo  {journal}
  {Phys. Rev. Res.}\ }\textbf {\bibinfo {volume} {2}},\ \bibinfo {pages}
  {013068} (\bibinfo {year} {2020})}\BibitemShut {NoStop}%
\bibitem [{\citenamefont {Chin}\ \emph {et~al.}(2010)\citenamefont {Chin},
  \citenamefont {Grimm}, \citenamefont {Julienne},\ and\ \citenamefont
  {Tiesinga}}]{chin_feshbach_2010}%
  \BibitemOpen
  \bibfield  {author} {\bibinfo {author} {\bibfnamefont {C.}~\bibnamefont
  {Chin}}, \bibinfo {author} {\bibfnamefont {R.}~\bibnamefont {Grimm}},
  \bibinfo {author} {\bibfnamefont {P.}~\bibnamefont {Julienne}}, \ and\
  \bibinfo {author} {\bibfnamefont {E.}~\bibnamefont {Tiesinga}},\ }\href
  {\doibase 10.1103/RevModPhys.82.1225} {\bibfield  {journal} {\bibinfo
  {journal} {Reviews of Modern Physics}\ }\textbf {\bibinfo {volume} {82}},\
  \bibinfo {pages} {1225} (\bibinfo {year} {2010})}\BibitemShut {NoStop}%
\end{thebibliography}%
\end{document}